\documentclass[preprint,showpacs,preprintnumbers,amsmath,amssymb]{revtex4}

\usepackage{graphicx}
\usepackage{dcolumn}
\usepackage{bm}

\begin{document}

\preprint{}

\title{Radial and Azimuthal Profiles of Two-Qubit/Rebit 
Hilbert-Schmidt Separability Probabilities and Related 3-D Visualization 
Analyses}

\author{Paul B. Slater}%
\email{slater@kitp.ucsb.edu}
\affiliation{%
ISBER, University of California, Santa Barbara, CA 93106\\
}%
\date{\today}

\begin{abstract}
{\it Firstly}, we reduce the long-standing problem of ascertaining the Hilbert-Schmidt probability
 that a generic pair of qubits is separable to that of determining
the specific nature of a one-dimensional (separability) function of the radial coordinate 
($r$) of the unit ball in 
15-dimensional Euclidean space, and similarly for a generic pair of rebits, using the 9-dimensional unit ball.  Separability probabilities, could, then,  be directly obtained by integrating the products of these functions (which we numerically estimate and plot) with jacobian factors of $r^{m}$  over $r \in [0,1]$, with $m=17$ for the two-rebit case, and $m=29$ in the two-qubit instance. {\it Secondly}, we repeat the analyses, but for the replacement of $r$ as the free variable, by the azimuthal angle 
$\phi \in [0,2 \pi]$--with the associated jacobian factors now being, trivially, unity.  So, the separability probabilities, then, become simply the areas under the curves generated. 
For our analyses, we employ 
an interesting  Cholesky-decomposition parameterization of the $4 \times 4$ density matrices. {\it Thirdly}, in an exploratory investigation, 
we examine the Hilbert-Schmidt separability probability question using the 
three-dimensional visualization (cube/tetrahedron/octahedron) 
of two qubits associated with
Avron, Bisker and Kenneth, and Leinaas, Myrheim and Ovrum, among others.
We find--in its two-rebit counterpart--that the Hilbert-Schmidt separability probability--the ratio of the measure assigned to the octahedron to that for the tetrahedron--is vanishingly small, in 
apparent conformity with observations of Caves, Fuchs and Rungta regarding {\it real} quantum mechanics. However, the ratio of the measure assigned to entangled states to that for potential entanglement witnesses (represented by points in the cube) is small, $\approx 0.005$, but seemingly finite.
We, then, begin an investigation of the full 15-dimensional two-qubit form of the question.
\end{abstract}


\maketitle
\section{Introduction}
A 2006 monograph entitled "Geometry of Quantum States: An Introduction to Quantum Entanglement" \cite{ingemarkarol} \cite{bzreview} had received by September 2011 (according to scholar.google) more than 460 citations in the scientific literature. We aspire, in this communication, to contribute--in two distinct sets of analyses (secs.~\ref{CD} and \ref{3D})--to this general area of strong current interest.
As in a number of previous studies  \cite{advances,ratios,maxconcur4,JMP2008,slater833,slaterJGP2,slaterPRA2}, we will be interested in determining (univariate) "separability functions", and with their use, "separability probabilities" \cite{ZHSL}. However, 
the analytical (Cholesky-decomposition) parameterization of density matrices which we, firstly, employ in sec.~\ref{CD}, is different than the (Euler-angle, Bloore/correlation) parameterizations earlier utilized, and our results here reflect different aspects of the one and the same, 
fundamental underlying, still not fully resolved, separability probability  problem.
\section{Cholesky-Decomposition Parameterization Analysis} \label{CD}
We begin by noting that in three preprints in the past decade, Ramakrishna and various colleagues
have examined a number of parameterizations of nonnegative-definite matrices (which include, of course, among them, the density matrices of 
quantum theory--bearing the additional requirement beyond nonnegative-definiteness of unit trace) \cite{constantinescu,tseng,tseng2}. One of these parameterizations--that will be the focus of our study here--is based on the well-known Cholesky-decomposition (cf. \cite{banaszek,daboul,ver2}),  which is  of important use in the numerical solution of linear equations, among other areas. 

We can represent a density matrix 
$\rho$, using the decomposition, as the product of a lower-triangular matrix 
$\Gamma=||\Gamma_{kj}||$--so that $\Gamma_{kj} =0$, $j>k$--with nonnegative entries ($\Gamma_{kk} \geq 0$) on the diagonal, and the transpose of its complex conjugate. Thus, 
\begin{equation}
\rho =\Gamma \Gamma^{\dagger}.
\end{equation}
It can be seen that the determinant of an $n \times n$  density matrix 
$\rho$, having imposed the unit trace requirement $\rho_{nn}=1-\Sigma_{k=1}^{n-1} \rho_{kk}$, takes the form (cf. \cite[p. 8]{tseng} for a {\it different} form of determinantal expression; also 
\cite[secs. D.2 and D.3]{slaterDunkl}),
\begin{equation} \label{CHOLESKY}
|\rho|= \Big( \Pi_{k=1}^{n-1} \Gamma_{kk} \Big)^2  
(1-\Sigma_{k=1}^{n-1} \Gamma_{kk}^2 -\Sigma_{k=1}^{n} \Sigma_{j<k}^{k-1}
|\Gamma_{kj}|^{2}).
\end{equation}
Further,
\begin{equation}
\rho_{nn}=(1-\Sigma_{k=1}^{n-1} \Gamma_{kk}^2 -\Sigma_{k=1}^{n-1} \Sigma_{j<k}^{k-1}
|\Gamma_{kj}|^{2})
\end{equation}
Thus, if the second factor  in (\ref{CHOLESKY}) is nonnegative, 
$\rho_{nn}$--which is at least as large as this factor--must also be nonnegative, and all 
the requirements that $\rho$ be a density matrix are fully satisfied.
One readily perceives, then, that for the case $n=4$, the two-qubit states ($\rho$) can be put into one-to-one correspondence with the points--imposing  the triad of restrictions $\Gamma_{kk} >0$ ($k=1,2,3$)-of {\it one-eighth} of the unit ball (centered at the origin, of course) in 15-dimensional Euclidean 
space.  Further, for the two-rebit states--that is those 
assigned $4 \times  4$ density matrices, the entries of which are restricted to real values  \cite{carl,batle2,slaterRebits}----the one-to-one correspondence is with the points of an analogous one-eighth part 
(a "hemidemisemi-hypersphere") of the unit ball  in 9-dimensional space. 

The jacobian for the transformation to the lower-triangular $\Gamma$ parameters 
is
\begin{equation} \label{jac1}
jac_{two-rebit} =  8 \Gamma_{11}^4 \Gamma_{22}^3 \Gamma_{33}^2 \end{equation}
 in the two-rebit case, and 
\begin{equation} \label{jac2}
jac_{two-qubit} = 8 \Gamma_{11}^7 \Gamma_{22}^5 \Gamma_{33}^3 \end{equation}
in the two-qubit case.

So we see, it is quite natural--given the Cholesky-decomposition parameterization and the sum-of-squares nature of the second factor in 
(\ref{CHOLESKY})--to convert the $\Gamma$'s to hyperspherical coordinates 
($r,\theta_{1},\theta_{2},\ldots$) (cf. \cite[eq. (A.2)]{daboul}). For the radial coordinate ($r$) we can then write (cf. (\ref{CHOLESKY})),
\begin{equation}
r=\sqrt{\Sigma_{k=1}^{n-1} \Gamma_{kk}^2 +\Sigma_{k=1}^{n} \Sigma_{j<k}^{k-1}
|\Gamma_{kj}|^{2}}
\end{equation}
 in these two differing ($n=9, 15$) dimensional settings. (Of course,  
$|\Gamma_{kj}|^2= \Gamma_{kj}^{2}$ in the two-rebit case.) 

We should, importantly, point out that the origin of the unit balls in both the two-qubit and two-rebit cases does {\it not} correspond, as one might initially expect, to the classical mixed state--having a diagonal density matrix with all entries equal to $\frac{1}{4}$--but to a [pure] diagonal density matrix with one entry equal to 1 and the other three to 0. By specifically choosing to set $\rho_{44}=1-\rho_{11}-\rho_{22}-\rho_{33}$, we have in our analyses that this unit diagonal-entry lies in the (4,4)-position.

Our principal objectives in this section of the study are to determine the relative prevalence--with respect to the Hilbert-Schmidt measure \cite[sec. 14.3]{ingemarkarol}  \cite{szHS,andai}--of separable {\it vs.} nonseparable/entangled states as: (1)  the radial coordinate $r$ of the unit ball 
varies between 0 (corresponding to the origin of the unit ball) and 1 
(its boundary) (sec.~\ref{RadialSec}); and (2) the azimuthal coordinate $\phi$ varies between 0 and $2 \phi$ (sec.~\ref{AzimuthalSec}). From the results, we will be able to directly (through {\it univariate} numerical integrations) obtain estimates of the probabilities of separability \cite{ZHSL}. We also aspire to discerning if the associated curves (separability functions) appear to assume any special functional form--in which case, the separability probabilities themselves
might be perceptible. 
\subsection{Radial Analyses} \label{RadialSec}
We can now make an important observation. 
The requirement that the partial transpose of $\rho$ have a nonnegative
 determinant--and, thus, corresponds to a separable state \cite{augusiak}--takes the general (quadratic)
form (in both the two-qubit and two-rebit cases),
\begin{equation}  \label{simpleform}
f(\theta_{1},\theta_{2},\ldots) - g(\theta_{1},\theta_{2},\ldots) r^2>0.
\end{equation}
Here, of course, the (apparently quite complicated) functions $f$ and $g$ of the angular (non-radial) coordinates are specific to the two-qubit and two-rebit scenarios. (Obviously, for 
$f=g=1$, (\ref{simpleform}) simply reduces back to the hyperspheres [unit balls] in question--one-eighth parts of which are, as already noted, the loci of the associated feasible $4 \times 4$ density matrices.)

If we enforce the separability requirement (\ref{simpleform}), while numerically integrating the corresponding jacobians over the angular coordinates (fourteen in number in the two-qubit case, and eight in the two-rebit case), and normalize the results by the corresponding Hilbert-Schmidt volumes \cite{szHS}--that is, $\frac{\pi^4}{967680}$ (two-rebit volume) and $\frac{\pi^6}{108972864000}$ (two-qubit volume)--we obtain the curves (radial profiles) displayed in Figs.~\ref{fig:realcurve} and \ref{fig:complexcurve}. 

The integrands used in this process are the  products of the indicated jacobians ((\ref{jac1}),(\ref{jac2})) with the jacobians  for the transformations 
from the lower-triangular  parameters  $\Gamma_{kj}$ to hyperspherical coordinates. (The exponent of $r$ in the resultant composite two-rebit jacobian is 17=9+8, while it is 29=15+14 in the two-qubit counterpart. The integers 8 and 14 are contributed by the standard jacobians for the transformations from cartesian to hyperspherical coordinates, and the 
integers $9 =4 +3 +2$ and $15=7+5+3$ from the jacobians ((\ref{jac1}),({\ref{jac2})) for the transformation to Cholesky-decomposition variables.) We employ numerical [Monte Carlo] integration--with 10,600,000 fourteen-dimensional points in the two-qubit case, and 1,100,000 eight-dimensional points in the two-rebit case. (More points were used in the former case, since due to the "curse-of-dimensionality", convergence to a smooth separability function was decidedly slower there. The use of quasi-Monte Carlo procedures \cite{giray1,tezuka}--which we will use for Figs.~\ref{fig:complexcurveazimuthal} and \ref{fig:visualization}--might also have been somewhat more computationally effective.) 

For each individual randomly-generated point, we evaluate whether the separability constraint (\ref{simpleform}) is satisfied or not at each of 10,001 equally-spaced values of $r \in [0,1]$. (We have also attempted to determine--by enforcing, through symbolic means, the separability constraint (\ref{simpleform}) with $r=0$--the exact theoretical values of the $y$-intercepts in the two figures, but have so far been unsuccessful in this quest. Clearly, the enforcement in symbolic terms of (\ref{simpleform}) with $r$ free to vary--that is, the separability probability question in its full 
generality--seems even more computationally-challenging.)

The two-rebit curve in Fig.~\ref{fig:realcurve}, if scaled to equal 1 at $r=0$,  can 
be very well-fitted, we found, by a curve (Figs.~\ref{fig:goodbetafit} and 
~\ref{fig:goodbetafit2})
\begin{equation} \label{BBETA}
1- I_{r}(r,3,\frac{1}{4})=1 - \frac{\int_{0}^{r} t^2 (1-t)^{-3/4}}{\int_{0}^{1} t^2 (1-t)^{-3/4}}= I_{1-r}(1-r,\frac{1}{4},3).
\end{equation}
Here $I_{r}(r,a,b)$ is a regularized incomplete beta function \cite{handbook}.

\begin{figure}
\includegraphics{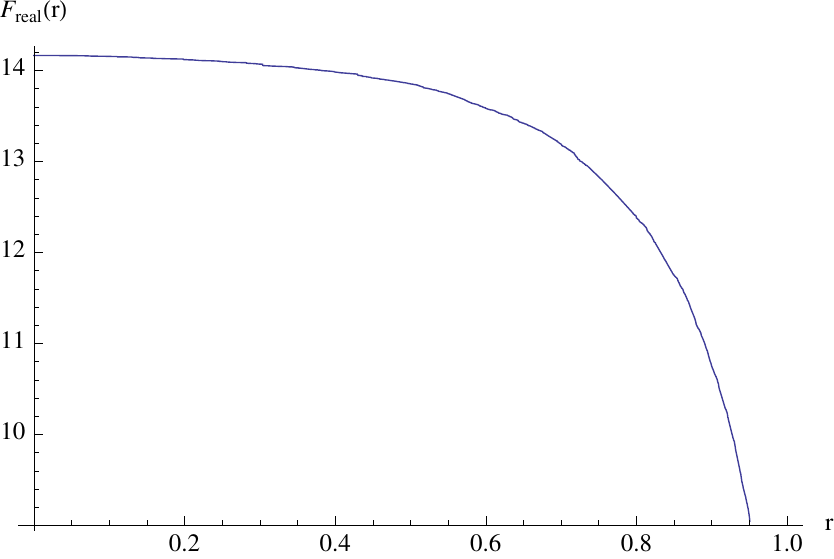}
\caption{\label{fig:realcurve}The relative prevalence--with respect to the Hilbert-Schmidt measure--of two-{\it rebit} separability, as a function of the radial coordinate $r$ of the unit ball in 8-dimensional Euclidean space. The two-rebit separability probability estimate--0.465885--is obtained by integrating the product of this curve and $r^{17}$ over 
$r \in [0,1]$. The exact probability conjectured 
in \cite{slater833} was $\frac{8}{17} \approx 0.470588$.}
\end{figure}
\begin{figure}
\includegraphics{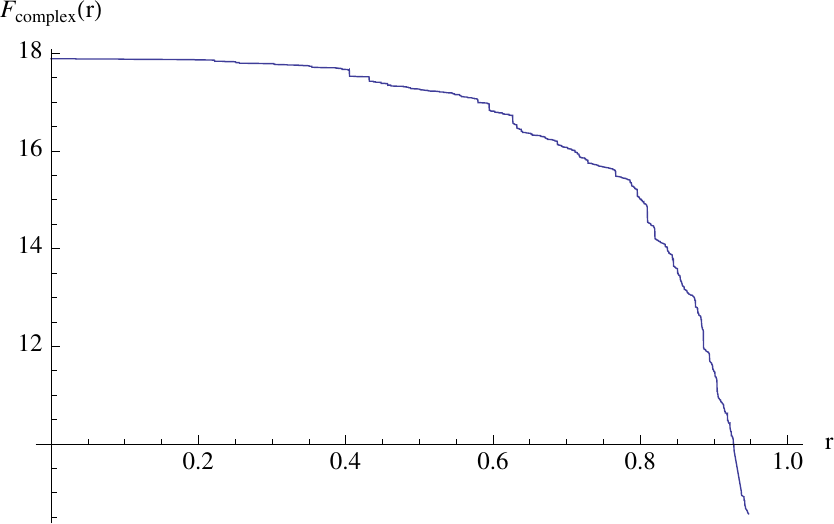}
\caption{\label{fig:complexcurve}The relative prevalence--with respect to the Hilbert-Schmidt measure--of two-{\it qubit} separability, as a function of the radial coordinate $r$ of the unit ball in 15-dimensional Euclidean space. The two-qubit separability probability estimate-0.209417--is obtained by integrating the product of this curve and $r^{29}$ over $r \in [0,1]$. The exact probability conjectured 
in \cite{slater833} was $\frac{8}{33} \approx 0.242424$.}
\end{figure}
\begin{figure}
\includegraphics{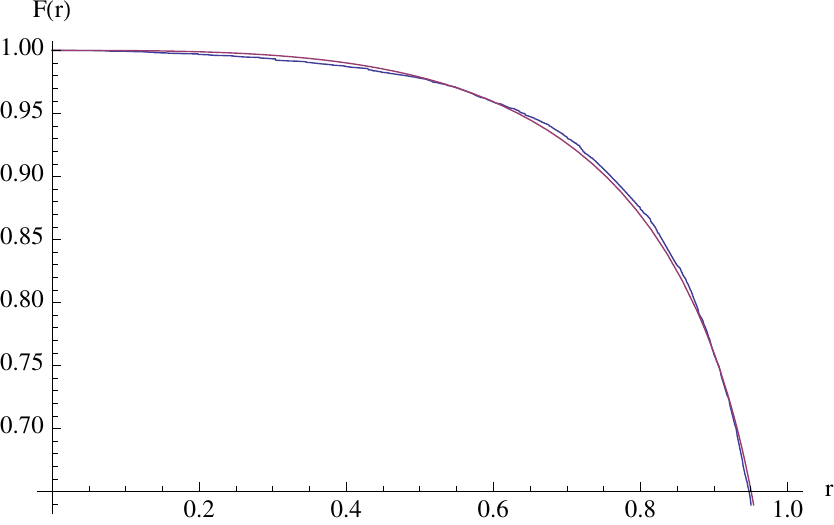}
\caption{\label{fig:goodbetafit}The (two-rebit) curve in Fig.~\ref{fig:realcurve}, scaled to equal 1 at $r=0$, plotted along with 
the closely-fitting function (\ref{BBETA})}
\end{figure}
\begin{figure}
\includegraphics{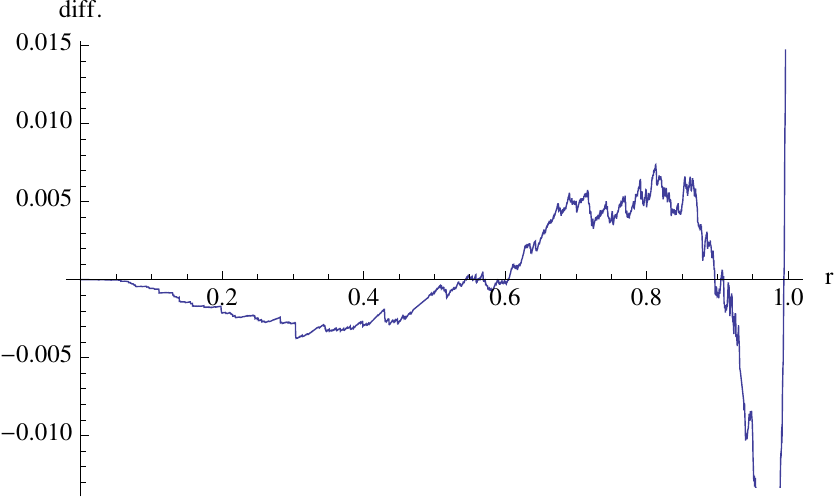}
\caption{\label{fig:goodbetafit2} The (two-rebit) curve in Fig.~\ref{fig:realcurve}, scaled to equal 1 at $r=0$, {\it minus} the 
fitted function (\ref{BBETA})}
\end{figure}

Let us jointly plot (Fig.~\ref{fig:combinedcurves}) 
the two-qubit curve (Fig.~\ref{fig:complexcurve}), along with the 
$\frac{3}{2}$-power of the two-rebit curve (Fig.~\ref{fig:realcurve}), after scaling  both curves to assume the value 1 at $r=0$. The rather close fit here is somewhat puzzling, as earlier studies of ours (reviewed, in detail, immediately below) of 
other "separability functions" would seem to suggest--in line with the 
"Dyson-index ansatz" of random matrix theory--that a power of 2 would be the appropriate one \cite{ratios,slater833}.
\begin{figure}
\includegraphics{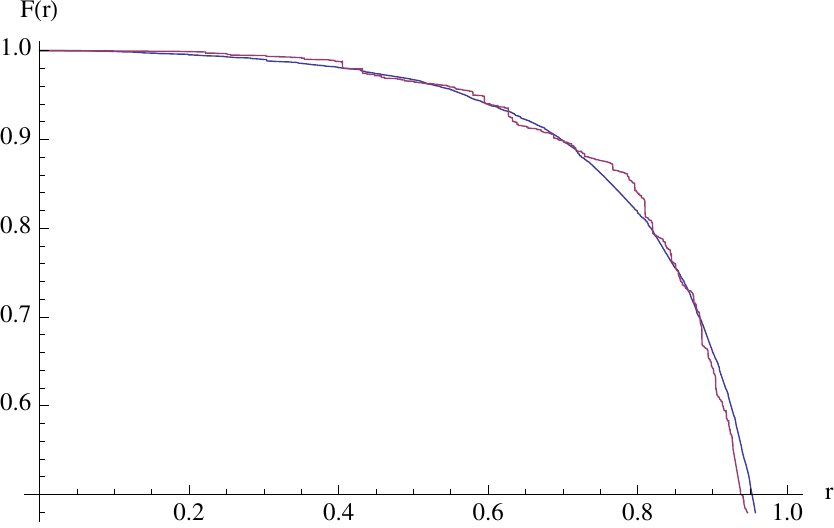}
\caption{\label{fig:combinedcurves}A joint plot of the two-qubit (rougher) curve (Fig.~\ref{fig:complexcurve}) and the $\frac{3}{2}$-power of the two-rebit curve (Fig.~\ref{fig:realcurve})--after scaling them both to equal 1 at $r=0$.}
\end{figure}
\subsubsection{Relations to earlier analyses}
The curves in our various figures might be termed "separability functions" of the radial coordinate $r$.
From such a perspective, they fulfill  a role strongly analogous to the (also univariate) separability functions estimated using 
density-matrix parameterizations 
(Euler angle, Bloore/correlation)--other than the Cholesky-decomposition one used above--in a number of related studies of ours.  \cite{advances,ratios,maxconcur4,JMP2008,slater833,slaterJGP2,slaterPRA2},

When employing the Bloore/correlation framework, the variable of interest (the counterpart to the radial coordinate $r$ above) was the 
diagonal-entry ratio (or its square root or logarithm) of 
$\nu=\frac{\rho_{11} \rho_{44}}{\rho_{22} \rho_{33}}$. When relying upon Euler angles \cite{tbs}, the variable of interest was the {\it maximal concurrence} \cite{ver,roland2}, that is, 
$C_{max}=\max{(\lambda_{1}-\lambda_3 -2 \sqrt{\lambda_2 \lambda_4)}}$, where the 
$\lambda$'s are the eigenvalues of $\rho$, ordered so that
$\lambda_1 \geq \lambda_2 \geq \lambda_3 \geq \lambda_4$. 
In the former (Bloore/correlation) set of analyses, the two-qubit separability function gave evidence of being proportional to the {\it square} of the two-rebit separability function.
In the latter (Euler-angle) set of analyses, there were--quite 
remarkably and still unexplainedly--pronounced jump discontinuities--in both the two-rebit and two-qubit cases--in the associated separability functions at $C_{max}=\frac{1}{2}$. For $C_{max} \in [\frac{1}{2},1]$, but not for  $C_{max} \in [0,\frac{1}{2}]$, the two functions appeared to adhere--as in the Bloore case, as just noted--to the Dyson-index ansatz 
($\beta=1, 2, 4$) of random matrix theory.

In the interest of completeness, in particular in light of the 
($\frac{3}{2}$)-power relationship apparently adhered to in Fig.~\ref{fig:combinedcurves}, let us review the results given in \cite{slater833} pertaining to separability functions parameterized 
(in the Bloore/correlation framework)  
by $\nu=\frac{\rho_{11} \rho_{44}}{\rho_{22} \rho_{33}}$.
The two-rebit jacobian based on the transformation to the $\nu$ variable of the Hilbert-Schmidt
volume element 
$(\rho_{11} \rho_{22} \rho_{33} \rho_{44})^{3/2}$ \cite{andai} took the form,
\begin{equation} \label{Jacreal}
\mathcal{J}_{real}(\nu) = \frac{\nu ^{3/2} \left(12 \left(\nu  (\nu +2) \left(\nu
   ^2+14 \nu +8\right)+1\right) \log \left(\sqrt{\nu
   }\right)-5 \left(5 \nu ^4+32 \nu ^3-32 \nu
   -5\right)\right)}{3780 (\nu -1)^9},
\end{equation}
while the separability function--based on numerical evidence--was surmised to be
\begin{equation} \label{newbeta}
I_{\nu}(\frac{1}{2},2) = \frac{1}{2} (3 -\nu) \sqrt{\nu}.
\end{equation}
One can perform the integration
\begin{equation} \label{almost1}
2 \int_{0}^{1} \mathcal{J}_{real}(\nu)  I_{\nu}(\nu,\frac{1}{2},2) d \nu = 
\frac{1}{151200} = \frac{1}{2^5 \cdot 3^3 \cdot 5^2 \cdot 7}, 
\end{equation}
and then obtain the conjecture made in \cite{slater833} for the Hilbert-Schmidt two-rebit separability probability, that is
\begin{equation} \label{almost2}
\frac{8}{17}=\frac{2419200}{17} \int_{0}^{1} \mathcal{J}_{real}(\nu)  I_{\nu}(\nu,\frac{1}{2},2) d \nu.
\end{equation}
Also, the further conjecture was made
for the Hilbert-Schmidt two-qubit separability probability 
(note--pursuant to the Dyson-index ansatz--involving
the {\it square} of the function),
\begin{equation} \label{almost3}
\frac{8}{33}=\frac{48432384000}{17} \int_{0}^{1} \mathcal{J}_{complex}(\nu)  I_{\nu}(\nu,\frac{1}{2},2)^2 d \nu. 
\end{equation}
Here $\mathcal{J}_{complex}(\nu)$ is the complex counterpart, the two-qubit jacobian based on the transformation to the $\nu$ variable of the Hilbert-Schmidt
volume element 
$(\rho_{11} \rho_{22} \rho_{33} \rho_{44})^{3}$ \cite{andai}--with a somewhat more intricate expression--to (\ref{Jacreal}).

In \cite{slaterRebits}, it was found in the generic two-rebit case that the determinants $|\rho|$ and $|\rho^{PT}|$ ("PT" denoting partial transposition)--see (\ref{det1}), (\ref{det2}) below--were {\it orthogonal} over the associated nine-dimensional space with respect to Hilbert-Schmidt measure. Now, having recognized 
(sec.~\ref{CD}) that this space can--using the Cholesky-decomposition parameterization--be viewed as a 
one-eighth part of the nine-dimensional unit ball 
(cf. \cite[eq. (A,2)]{daboul}), it may be 
more readily possible to classify these multivariate orthogonal polynomials (MOPS) \cite{dumitriu,dunkl2} into standard forms (cf. \cite{griffithsspano}).

\subsection{Azimuthal Analyses} \label{AzimuthalSec}
Now, let us repeat our two-qubit and two-rebit analyses above, but now using--instead of the radial coordinate $r$--as our free variable, the {\it azimuthal} angle ($\phi$) in each of the two hyperspherical coordinate systems. (The counterparts of the separability requirement (\ref{simpleform}) now take the somewhat more complicated forms of  quadratic  polynomials in $\sin{\phi}$ and $\cos{\phi}$.)
Then, $\phi$ is the only angle with a range $[0, 2\pi]$ (rather than 
$[0,\pi]$), and also the only angle {\it absent} from the associated jacobians over which the multidimensional integrations must be performed. 

Therefore, to obtain separability probability estimates from the associated   "azimuthal profiles" we will only need to integrate the profiles of 
$\hat{\phi} =\frac{\phi}{2 \pi}$ over 
$\hat{\phi} \in  [0, 1]$ 
without multiplication by any (non-trivial) jacobian factor at all 
(which were $r^{17}$ and $r^{29}$ in the radial coordinate case). 
(For each sample point generated, we now evaluate the determinant at each of 1,001 equally-spaced values of $\hat{\phi}$, rather than the 10,001 equally-spaced values of $r$ used in the earlier set of analyses. For the two-rebit case, 22,825,000 eight-dimensonal Monte Carlo-generated points 
were employed, while for the two-qubit case, 17,460,000 fourteen-dimensional {\it quasi}-Monte Carlo points were used--given that simple Monte Carlo results in this latter case seemed highly unstable in nature. 
Nevertheless, even with this change in computational strategy, considerable instability appears to remain.) 

Thus, the  separability probability estimates would simply be the areas under the
$\hat{\phi}$-parameterized separability functions (Figs.~\ref{fig:realcurveazimuthal} and \ref{fig:complexcurveazimuthal})-if the 
horizontal axes were drawn to intercept at $F(\hat{\phi})=0$. 
(At such a resolution, the two curves look largely {\it flat} in nature.)
In the two-rebit case, there appears to be oscillatory behavior symmetrical 
(perhaps sinusoidal in nature) around  $\hat{\phi}=\frac{1}{2}$ (or $\phi= \pi$) (Fig.~\ref{fig:realcurveazimuthalfit}). In fact, it seems possible to consider Fig.~\ref{fig:realcurveazimuthal} as consisting of a single {\it flat} baseline, with the oscillations around the baseline simply averaging out to zero. In such a case, it would be irrelevant to the chief goal of determining the HS separability probability, what the specific amplitude of these 
({\it zero}-average) oscillations about the determinative baseline is.
Fig.~\ref{fig:complexcurveazimuthal} serves as the two-qubit analogue
of  Fig.~\ref{fig:realcurveazimuthal}.
\begin{figure}
\includegraphics{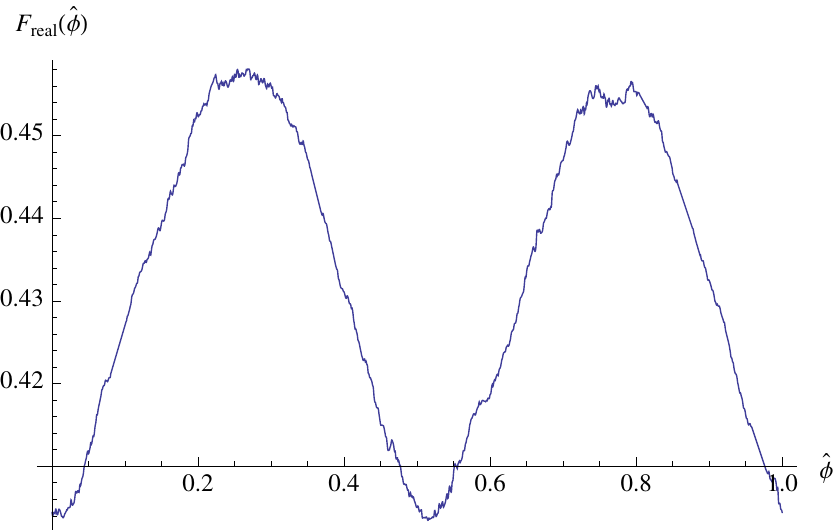}
\caption{\label{fig:realcurveazimuthal}The relative prevalence--with respect to the Hilbert-Schmidt measure--of two-rebit separability, as a function of the azimuthal coordinate $\hat{\phi} =\frac{\phi}{2 \pi}$ of the unit ball in 8-dimensional Euclidean space. The two-rebit separability probability estimate--0.433082--is obtained by integrating  this curve 
(the jacobian factor is trivially unity) over $\hat{\phi} \in [0,1]$. The exact probability conjectured 
in \cite{slater833} was $\frac{8}{17} \approx 0.470588$.}
\end{figure}
\begin{figure}
\includegraphics{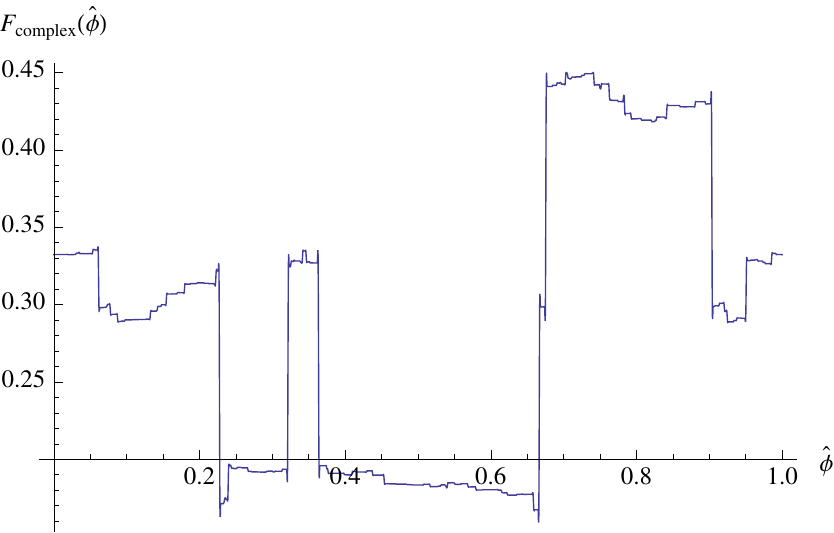}
\caption{\label{fig:complexcurveazimuthal}The relative prevalence--with respect to the Hilbert-Schmidt measure--of two-qubit separability, as a function of the azimuthal coordinate $\hat{\phi}$ of the unit ball in 15-dimensional Euclidean space. The two-qubit separability probability estimate--0.289656--is obtained by integrating  this curve 
(the jacobian factor is trivially unity) over $\hat{\phi} \in [0,1]$. The exact probability conjectured 
in \cite{slater833} was $\frac{8}{33} \approx 0.242424$.}
\end{figure}
\begin{figure}
\includegraphics{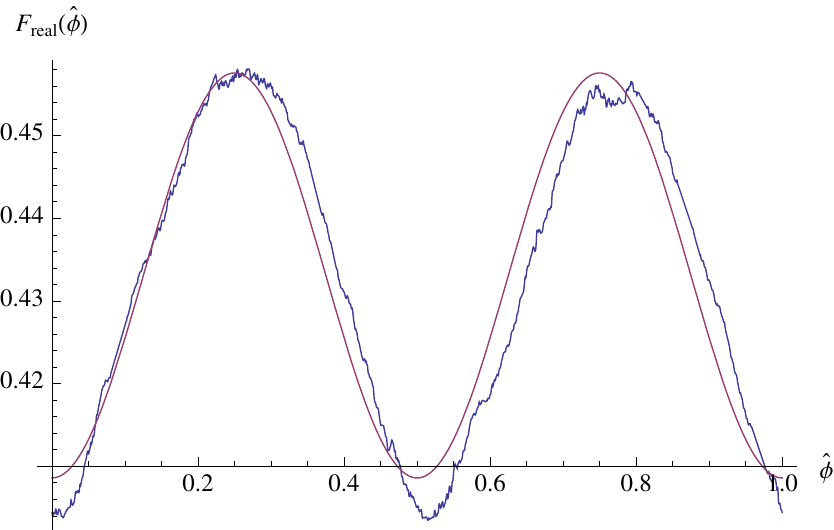}
\caption{\label{fig:realcurveazimuthalfit}A fit to the curve 
(Fig.~\ref{fig:realcurveazimuthal}) by the oscillatory function
$0.433082+0.0244966  \cos{4 \pi \hat{\phi}}$}
\end{figure}

We have, above, used as free variables, the radial and azimuthal coordinates. Of course, we might, similarly examine any or all 
of the remaining
(non-azimuthal, polar) angles (13 [two-qubit] or 7 [two-rebit] ones). In the four distinct (two radial in nature, and two, azimuthal, as well as two two-rebit and two two-qubit) Monte-Carlo 
and quasi-Monte Carlo analyses, reported above, we would certainly have wished to have been able to employ even more points than we have, in the face of computational limitations, been able to. (In any case, the computational resources employed--a number of months on MacMini machines--have been quite extensive.)

\section{Three-Dimensional Visualization-Based Analyses} \label{3D}
Another interesting, related topic we have investigated is the possibility of addressing the Hilbert-Schmidt separability probability problem within the 
"visualization" (cube/tetrahedron/octahedron) framework of Avron, Bisker and Kenneth \cite{avron,avron2} (cf. \cite[eq. (1)]{ver8} 
\cite[Fig. 4]{leinaas}, also \cite{SWZ,SWZ2,langcaves}).
("The octahedron represents the equivalence class of separable states. 
The set of points that lie outside the octahedron but inside the tetrahedron represent the equivalence class of entangled states. The set of points that lie outside the tetrahedron but inside the cube represent entanglement witnesses. The vertices of the tetrahedron represent the equivalence class of pure states. Points related by the tetrahedral symmetry represent the same equivalence class" \cite{avron2}.)

In this regard, we present Fig.~\ref{fig:visualization}, based on 128,000 
six-dimensional quasi-Monte Carlo points (for each such point, we compute the Hilbert-Schmidt measure on a $51 \times 51$ uniform grid on the $d_1-d_3$-plane--omitting the $d_2$-axis, since the  measure proves to be independent of $d_2$) parameterizing the tensor product
$A \otimes B$ of equation (1) of \cite{ver8}
\begin{equation} \label{rho16}
\rho'=\frac{(A \otimes B) \rho (A \otimes B)^{\dagger}}{\mbox{Tr}[(A \otimes B) \rho (A \otimes B)^{\dagger}]},
\end{equation} 
where ($\sigma_k$ is a Pauli matrix), and 
\begin{equation} \label{Paulis}
\rho=\frac{1}{4} \Big(I + \Sigma_{k=1}^3 d_i \sigma_k \otimes \sigma_k \Big), d_i \in [-1,1].
\end{equation} 
(The $2 \times 2$ three-parameter [unit determinant] matrices $A$ and $B$ belong to the special linear group $SL(2,R)$. The group itself has infinite 
volume \cite[App.  2]{simontaylor}.)

The octahedron, tetrahedron 
and cube are, of course, three-dimensional objects. But we have found that the Hilbert-Schmidt measure--in the generic two-rebit case--is constant across one of the three-dimensions
($d_2$ in the notation of \cite[sec. IV,B]{leinaas}). (Certainly, it is rather obvious--in partial  retrospect--that if $d_2 \neq 0$ in 
(\ref{Paulis}), then $\rho^{'}$ must have complex entries, and thus does not represent a two-rebit system.) So, we 
conveniently--for visual purposes--require only two dimensions ($d_1, d_3$) for the plot. For our parameterization of elements of the special linear group $SL(2,R)$ we 
employ a certain formula 
\cite[eqs. (1.37) and (1.38), p. 85]{miller}--drawn from 
Bargmann \cite{bargmann}--because for numerical purposes, all three of the parameters have finite range.
\begin{figure}
\includegraphics{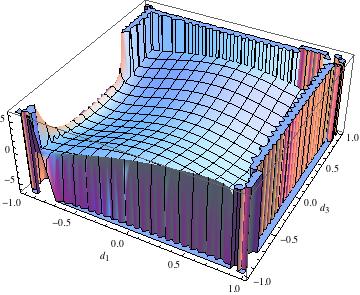}
\caption{\label{fig:visualization}The estimated Hilbert-Schmidt measure over the $d_1-d_3$-plane (cf. Fig. 4 of \cite{leinaas} and Figs. 1 of 
both \cite{avron,avron2}) of the three-dimensional cube of equivalence classes of two rebits. The measure is constant across the omitted
coordinate $d_2$. The associated separability probability--in conformity
to {\it real} quantum mechanics \cite{carl}--is zero.}
\end{figure}

The exceptional feature here--in contrast to our other investigations of "separability functions"--is that there is only one germane function--not two distinct ones (one for all states, and one just for separable states)--over the smaller number of parameters that is the domain of the separability function.

In this regard, we note that (cf. (\ref{rho16}))
\begin{equation} \label{det1}
|\rho^{'}|=\frac{\left(d_1-d_2-d_3-1\right) \left(d_1+d_2-d_3+1\right)
   \left(d_1-d_2+d_3+1\right) \left(d_1+d_2+d_3-1\right)}{256 (\mbox{Tr}[(A \otimes B) \rho (A \otimes B)^{\dagger}])^4}
\end{equation}
and, for the partial transpose of $\rho^{'}$, the slightly different expression (note the sign of 1 in the last factor of the two 
determinantal expressions)
\begin{equation} \label{det2}
|(\rho^{'})^{PT}|=\frac{\left(d_1-d_2-d_3+1\right) \left(d_1+d_2-d_3-1\right)
   \left(d_1-d_2+d_3-1\right) \left(d_1+d_2+d_3+1\right))}{256 (\mbox{Tr}[(A \otimes B) \rho (A \otimes B)^{\dagger}])^4}.
\end{equation}
(As already noted, these two determinants have been found to be mutually orthogonal with
respect to Hilbert-Schmidt measure \cite{slaterRebits}.)
Since the denominators of these last two determinantal expressions are necessarily positive, we see that the positivity of the determinants themselves depends only upon the  
numerators--which, importantly and exceptionally, are functions only of the three $d_i$'s, and not of any of the six $SL(2,R)$ parameters (for 
the $2 \times 2$ matrices $A$ and $B$).
\subsection{Significance of {\it real} quantum mechanics}
Our estimate--based on the results used in 
Fig.~\ref{fig:visualization}--of the ratio of the Hilbert-Schmidt measure assigned to the entire cube representing the equivalence classes of potential entanglement witnessess to the tetrahedron of all possible (entangled and non-entangled) states is 0.00584127. However, the estimates of the separability probability (the ratio of the Riemannian volume of the octahedron to that of the tetrahedron)--to our initial considerable surprise--was not on the order
of 0.45 as we have repeatedly observed, as noted above, in prior studies. We now found this probability to be vanishingly small. Nevertheless, these two disparate outcomes do seem reconcilable using a perspective provided by Caves, Fuchs and Rungta \cite{carl} (cf. \cite{batleplastino1}).

The crucial observation of these three authors--in this regard--is given in the penultimate paragraph of \cite{carl}:
\begin{quotation}
Another interesting fact is how the regions of entangled vs. separable
states within the full set of quantum states differ in going from
real to complex quantum mechanics. In the complex theory, the maximally 
mixed state $\rho=\frac{1}{4} I_4$ of two qubits is surrounded by an open 
set of separable states \ldots.  In the real 
theory, however, the states 
\begin{equation}
\rho^{AB}=\frac{1}{4} \Big(I \otimes I +\alpha (\sigma_y \otimes \sigma_y) \Big), \alpha \in [0,1]
\end{equation}
demonstrate that there 
are entangled states arbitrarily close to the maximally mixed state. 
This conclusion also follows from the fact that the condition for 
{\it real} (emphasis added) separability, $\mbox{tr} (\rho^{AB} \sigma_y 
\otimes \sigma_y)=0$, implies that the space of {\it real} (emphasis 
added) separable states is an 8-dimensional submanifold of the 9-dimensional manifold of all states.
\end{quotation}

Thus, it can be seen that our earlier estimates of a Hilbert-Schmidt separability probability of a pair of rebits on the order of 0.45 have been (quite legitimately, it would seem) obtained in the framework of {\it complex} quantum mechanics, while the vanishingly small probability observed in the analysis here, relying upon the two-qubit visualization scheme \cite{avron,avron2,ver8,leinaas}}, is a result of {\it real} quantum mechanics--since an 8-dimensional submanifold must necessarily have zero 
9-dimensional volume.

\begin{figure}
\includegraphics{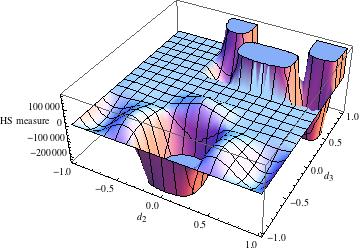}
\caption{\label{fig:Binnedd1}Estimated Hilbert-Schmidt measure over the 
$d_2-d_3$-plane for two-qubit systems}
\end{figure}

\begin{figure}
\includegraphics{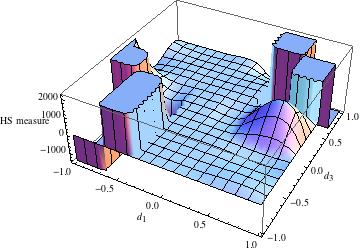}
\caption{\label{fig:Binnedd2}Estimated Hilbert-Schmidt measure over the 
$d_1-d_3$-plane for two-qubit systems}
\end{figure}

\begin{figure}
\includegraphics{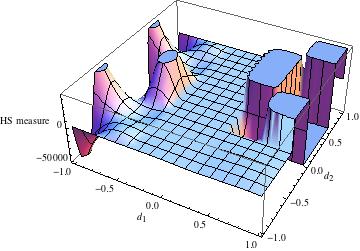}
\caption{\label{fig:Binnedd3}Estimated Hilbert-Schmidt measure over the 
$d_1-d_2$-plane for two-qubit systems}
\end{figure}

Of course, it would be of obvious interest--but computationally quite challenging--to address these 
measure-theoretic questions in 
the full two-qubit (rather than the somewhat degenerate two-rebit specialization) framework. 
We are presently investigating this matter.
In fact, let us present here some very preliminary results in this 
regard. We have employed 114,000 15-dimensional quasi-Monte Carlo points, parameterizing the two (six-dimensional) general linear groups employed in (\ref{rho16}), as well as the three variables $d_1, d_2, d_3$. (We allow the entries of the $2 \times 2$ matrices $A$ and $B$ to vary uniformly between -500 and 500.) We, further, bin the values of $d_, d_2, d_3 \in [-1,1]$ generated into $343=7^3$ uniformaly-sized bins. For each of these 15-dimensional points, we calculate the corresponding $15 \times 15$ jacobian determinant. (An interesting, importrant question that arises is under what conditions this determinant is {\it negative} (cf. \cite{kwokchen}).) The estimate of the Hilbert-Schmidt two-qubit separability probability obtained is 0.07954. Summing the bins over each of the three coordinates $d_i$, in turn, and interpolating the results over the $[-1,1]^2$ square, we obtain the three two-dimensional (marginal) surfaces 
depicted in 
Figs.~\ref{fig:Binnedd1}, 
\ref{fig:Binnedd2} and \ref{fig:Binnedd3}.
We will continue to add quasi-Monte Carlo points to these analyses, as well
as conduct parallel analyses with more refined bin sizes than above.

\begin{acknowledgments}
I would like to express appreciation to the Kavli Institute for Theoretical
Physics (KITP)
for computational support in this research, as well as to Joseph Avron and Oded Kenneth, and Sergio Cacciatori for relevant communications.
\end{acknowledgments}

\bibliography{Cholesky10}

\begin{thebibliography}{44}
\expandafter\ifx\csname natexlab\endcsname\relax\def\natexlab#1{#1}\fi
\expandafter\ifx\csname bibnamefont\endcsname\relax
  \def\bibnamefont#1{#1}\fi
\expandafter\ifx\csname bibfnamefont\endcsname\relax
  \def\bibfnamefont#1{#1}\fi
\expandafter\ifx\csname citenamefont\endcsname\relax
  \def\citenamefont#1{#1}\fi
\expandafter\ifx\csname url\endcsname\relax
  \def\url#1{\texttt{#1}}\fi
\expandafter\ifx\csname urlprefix\endcsname\relax\def\urlprefix{URL }\fi
\providecommand{\bibinfo}[2]{#2}
\providecommand{\eprint}[2][]{\url{#2}}

\bibitem[{\citenamefont{Bengtsson and {\.Z}yczkowski}(2006)}]{ingemarkarol}
\bibinfo{author}{\bibfnamefont{I.}~\bibnamefont{Bengtsson}} \bibnamefont{and}
  \bibinfo{author}{\bibfnamefont{K.}~\bibnamefont{{\.Z}yczkowski}},
  \emph{\bibinfo{title}{Geometry of Quantum States}}
  (\bibinfo{publisher}{Cambridge}, \bibinfo{address}{Cambridge},
  \bibinfo{year}{2006}).

\bibitem[{\citenamefont{Slater}(2007{\natexlab{a}})}]{bzreview}
\bibinfo{author}{\bibfnamefont{P.~B.} \bibnamefont{Slater}},
  \bibinfo{journal}{MathSciNet} p. \bibinfo{pages}{MR2230995 (2007k:81001)}
  (\bibinfo{year}{2007}{\natexlab{a}}).

\bibitem[{\citenamefont{Slater}(2010)}]{advances}
\bibinfo{author}{\bibfnamefont{P.~B.} \bibnamefont{Slater}},
  \bibinfo{journal}{J. Phys. A} \textbf{\bibinfo{volume}{43}},
  \bibinfo{pages}{195302} (\bibinfo{year}{2010}).

\bibitem[{\citenamefont{Slater}(2009{\natexlab{a}})}]{ratios}
\bibinfo{author}{\bibfnamefont{P.~B.} \bibnamefont{Slater}},
  \bibinfo{journal}{J. Phys. A} \textbf{\bibinfo{volume}{42}},
  \bibinfo{pages}{465305} (\bibinfo{year}{2009}{\natexlab{a}}).

\bibitem[{\citenamefont{Slater}(2008{\natexlab{a}})}]{maxconcur4}
\bibinfo{author}{\bibfnamefont{P.~B.} \bibnamefont{Slater}},
  \bibinfo{journal}{J. Phys. A} \textbf{\bibinfo{volume}{41}},
  \bibinfo{pages}{505303} (\bibinfo{year}{2008}{\natexlab{a}}).

\bibitem[{\citenamefont{Slater}(2009{\natexlab{b}})}]{JMP2008}
\bibinfo{author}{\bibfnamefont{P.~B.} \bibnamefont{Slater}},
  \bibinfo{journal}{J. Geom. Phys.} \textbf{\bibinfo{volume}{59}},
  \bibinfo{pages}{17} (\bibinfo{year}{2009}{\natexlab{b}}).

\bibitem[{\citenamefont{Slater}(2007{\natexlab{b}})}]{slater833}
\bibinfo{author}{\bibfnamefont{P.~B.} \bibnamefont{Slater}},
  \bibinfo{journal}{J. Phys. A} \textbf{\bibinfo{volume}{40}},
  \bibinfo{pages}{14279} (\bibinfo{year}{2007}{\natexlab{b}}).

\bibitem[{\citenamefont{Slater}(2008{\natexlab{b}})}]{slaterJGP2}
\bibinfo{author}{\bibfnamefont{P.~B.} \bibnamefont{Slater}},
  \bibinfo{journal}{J. Geom. Phys.} \textbf{\bibinfo{volume}{58}},
  \bibinfo{pages}{1101} (\bibinfo{year}{2008}{\natexlab{b}}).

\bibitem[{\citenamefont{Slater}(2007{\natexlab{c}})}]{slaterPRA2}
\bibinfo{author}{\bibfnamefont{P.~B.} \bibnamefont{Slater}},
  \bibinfo{journal}{Phys. Rev. A} \textbf{\bibinfo{volume}{75}},
  \bibinfo{pages}{032326} (\bibinfo{year}{2007}{\natexlab{c}}).

\bibitem[{\citenamefont{{\.Z}yczkowski
  et~al.}(1998)\citenamefont{{\.Z}yczkowski, Horodecki, Sanpera, and
  Lewenstein}}]{ZHSL}
\bibinfo{author}{\bibfnamefont{K.}~\bibnamefont{{\.Z}yczkowski}},
  \bibinfo{author}{\bibfnamefont{P.}~\bibnamefont{Horodecki}},
  \bibinfo{author}{\bibfnamefont{A.}~\bibnamefont{Sanpera}}, \bibnamefont{and}
  \bibinfo{author}{\bibfnamefont{M.}~\bibnamefont{Lewenstein}},
  \bibinfo{journal}{Phys. Rev. A} \textbf{\bibinfo{volume}{58}},
  \bibinfo{pages}{883} (\bibinfo{year}{1998}).

\bibitem[{\citenamefont{Constantinescu and Ramakrishna}()}]{constantinescu}
\bibinfo{author}{\bibfnamefont{T.}~\bibnamefont{Constantinescu}}
  \bibnamefont{and}
  \bibinfo{author}{\bibfnamefont{V.}~\bibnamefont{Ramakrishna}},
  \eprint{quant-ph/0306167}.

\bibitem[{\citenamefont{Tseng et~al.}()\citenamefont{Tseng, Zhou, and
  Ramakrishna}}]{tseng}
\bibinfo{author}{\bibfnamefont{M.~C.} \bibnamefont{Tseng}},
  \bibinfo{author}{\bibfnamefont{H.}~\bibnamefont{Zhou}}, \bibnamefont{and}
  \bibinfo{author}{\bibfnamefont{V.}~\bibnamefont{Ramakrishna}},
  \eprint{quant-ph/0610020}.

\bibitem[{\citenamefont{Tseng and Ramakrishna}()}]{tseng2}
\bibinfo{author}{\bibfnamefont{M.~C.} \bibnamefont{Tseng}} \bibnamefont{and}
  \bibinfo{author}{\bibfnamefont{V.}~\bibnamefont{Ramakrishna}},
  \eprint{quant-ph/0610021}.

\bibitem[{\citenamefont{Banaszek et~al.}(1999)\citenamefont{Banaszek, D'Ariano,
  Paris, and Sacchi}}]{banaszek}
\bibinfo{author}{\bibfnamefont{K.}~\bibnamefont{Banaszek}},
  \bibinfo{author}{\bibfnamefont{G.~M.} \bibnamefont{D'Ariano}},
  \bibinfo{author}{\bibfnamefont{M.~G.~A.} \bibnamefont{Paris}},
  \bibnamefont{and} \bibinfo{author}{\bibfnamefont{M.~F.}
  \bibnamefont{Sacchi}}, \bibinfo{journal}{Phys. Rev. A}
  \textbf{\bibinfo{volume}{61}}, \bibinfo{pages}{010304}
  (\bibinfo{year}{1999}).

\bibitem[{\citenamefont{Daboul}(1967)}]{daboul}
\bibinfo{author}{\bibfnamefont{J.}~\bibnamefont{Daboul}},
  \bibinfo{journal}{Nucl. Phys.} \textbf{\bibinfo{volume}{B4}},
  \bibinfo{pages}{180} (\bibinfo{year}{1967}).

\bibitem[{\citenamefont{Verstraete
  et~al.}(2001{\natexlab{a}})\citenamefont{Verstraete, Audenaert, Dehaene, and
  Moor}}]{ver2}
\bibinfo{author}{\bibfnamefont{F.}~\bibnamefont{Verstraete}},
  \bibinfo{author}{\bibfnamefont{K.}~\bibnamefont{Audenaert}},
  \bibinfo{author}{\bibfnamefont{J.}~\bibnamefont{Dehaene}}, \bibnamefont{and}
  \bibinfo{author}{\bibfnamefont{B.~D.} \bibnamefont{Moor}},
  \bibinfo{journal}{J. Phys. A} \textbf{\bibinfo{volume}{34}},
  \bibinfo{pages}{10327} (\bibinfo{year}{2001}{\natexlab{a}}).

\bibitem[{\citenamefont{Slater}({\natexlab{a}})}]{slaterDunkl}
\bibinfo{author}{\bibfnamefont{P.~B.} \bibnamefont{Slater}},
  \eprint{arXiv:1109.2560}.

\bibitem[{\citenamefont{Caves et~al.}(2001)\citenamefont{Caves, Fuchs, and
  Rungta}}]{carl}
\bibinfo{author}{\bibfnamefont{C.~M.} \bibnamefont{Caves}},
  \bibinfo{author}{\bibfnamefont{C.~A.} \bibnamefont{Fuchs}}, \bibnamefont{and}
  \bibinfo{author}{\bibfnamefont{P.}~\bibnamefont{Rungta}},
  \bibinfo{journal}{Found. Phys. Letts.} \textbf{\bibinfo{volume}{14}},
  \bibinfo{pages}{199} (\bibinfo{year}{2001}).

\bibitem[{\citenamefont{Batle et~al.}(2003)\citenamefont{Batle, Plastino,
  Casas, and Plastino}}]{batle2}
\bibinfo{author}{\bibfnamefont{J.}~\bibnamefont{Batle}},
  \bibinfo{author}{\bibfnamefont{A.~R.} \bibnamefont{Plastino}},
  \bibinfo{author}{\bibfnamefont{M.}~\bibnamefont{Casas}}, \bibnamefont{and}
  \bibinfo{author}{\bibfnamefont{A.}~\bibnamefont{Plastino}},
  \bibinfo{journal}{Opt. Spect.} \textbf{\bibinfo{volume}{94}},
  \bibinfo{pages}{1562} (\bibinfo{year}{2003}).

\bibitem[{\citenamefont{Slater}({\natexlab{b}})}]{slaterRebits}
\bibinfo{author}{\bibfnamefont{P.~B.} \bibnamefont{Slater}},
  \eprint{arXiv:1007.4805}.

\bibitem[{\citenamefont{{\.Z}yczkowski and Sommers}(2003)}]{szHS}
\bibinfo{author}{\bibfnamefont{K.}~\bibnamefont{{\.Z}yczkowski}}
  \bibnamefont{and} \bibinfo{author}{\bibfnamefont{H.-J.}
  \bibnamefont{Sommers}}, \bibinfo{journal}{J. Phys. A}
  \textbf{\bibinfo{volume}{36}}, \bibinfo{pages}{10115} (\bibinfo{year}{2003}).

\bibitem[{\citenamefont{Andai}(2006)}]{andai}
\bibinfo{author}{\bibfnamefont{A.}~\bibnamefont{Andai}}, \bibinfo{journal}{J.
  Phys. A} \textbf{\bibinfo{volume}{39}}, \bibinfo{pages}{13641}
  (\bibinfo{year}{2006}).

\bibitem[{\citenamefont{Augusiak et~al.}(2008)\citenamefont{Augusiak,
  Horodecki, and Demianowicz}}]{augusiak}
\bibinfo{author}{\bibfnamefont{R.}~\bibnamefont{Augusiak}},
  \bibinfo{author}{\bibfnamefont{R.}~\bibnamefont{Horodecki}},
  \bibnamefont{and}
  \bibinfo{author}{\bibfnamefont{M.}~\bibnamefont{Demianowicz}},
  \bibinfo{journal}{Phys. Rev.} \textbf{\bibinfo{volume}{77}},
  \bibinfo{pages}{030301(R)} (\bibinfo{year}{2008}).

\bibitem[{\citenamefont{{\"O}kten}(1999)}]{giray1}
\bibinfo{author}{\bibfnamefont{G.}~\bibnamefont{{\"O}kten}},
  \bibinfo{journal}{MATHEMATICA in Educ. Res.} \textbf{\bibinfo{volume}{8}},
  \bibinfo{pages}{52} (\bibinfo{year}{1999}).

\bibitem[{\citenamefont{Faure and Tezuka}(2002)}]{tezuka}
\bibinfo{author}{\bibfnamefont{H.}~\bibnamefont{Faure}} \bibnamefont{and}
  \bibinfo{author}{\bibfnamefont{S.}~\bibnamefont{Tezuka}}, in
  \emph{\bibinfo{booktitle}{Monte Carlo and Quasi-Monte Carlo Methods 2000
  (Hong Kong)}}, edited by \bibinfo{editor}{\bibfnamefont{K.~T.}
  \bibnamefont{Tang}}, \bibinfo{editor}{\bibfnamefont{F.~J.}
  \bibnamefont{Hickernell}}, \bibnamefont{and}
  \bibinfo{editor}{\bibfnamefont{H.}~\bibnamefont{Niederreiter}}
  (\bibinfo{publisher}{Springer}, \bibinfo{address}{Berlin},
  \bibinfo{year}{2002}), p. \bibinfo{pages}{242}.

\bibitem[{\citenamefont{Gupta and Nadarajah}(2004)}]{handbook}
\bibinfo{author}{\bibfnamefont{A.~K.} \bibnamefont{Gupta}} \bibnamefont{and}
  \bibinfo{author}{\bibfnamefont{S.}~\bibnamefont{Nadarajah}},
  \emph{\bibinfo{title}{Handbook of Beta Distribution and Its Applications}}
  (\bibinfo{publisher}{Marcel Dekker}, \bibinfo{address}{New York},
  \bibinfo{year}{2004}).

\bibitem[{\citenamefont{Tilma et~al.}(2002)\citenamefont{Tilma, Byrd, and
  Sudarshan}}]{tbs}
\bibinfo{author}{\bibfnamefont{T.}~\bibnamefont{Tilma}},
  \bibinfo{author}{\bibfnamefont{M.}~\bibnamefont{Byrd}}, \bibnamefont{and}
  \bibinfo{author}{\bibfnamefont{E.~C.~G.} \bibnamefont{Sudarshan}},
  \bibinfo{journal}{J. Phys. A} \textbf{\bibinfo{volume}{35}},
  \bibinfo{pages}{10445} (\bibinfo{year}{2002}).

\bibitem[{\citenamefont{Verstraete
  et~al.}(2001{\natexlab{b}})\citenamefont{Verstraete, Audenaert, and
  DeMoor}}]{ver}
\bibinfo{author}{\bibfnamefont{F.}~\bibnamefont{Verstraete}},
  \bibinfo{author}{\bibfnamefont{K.}~\bibnamefont{Audenaert}},
  \bibnamefont{and} \bibinfo{author}{\bibfnamefont{B.}~\bibnamefont{DeMoor}},
  \bibinfo{journal}{Phys. Rev. A} \textbf{\bibinfo{volume}{64}},
  \bibinfo{pages}{012316} (\bibinfo{year}{2001}{\natexlab{b}}).

\bibitem[{\citenamefont{Hildebrand}(2007)}]{roland2}
\bibinfo{author}{\bibfnamefont{R.}~\bibnamefont{Hildebrand}},
  \bibinfo{journal}{J. Math. Phys.} \textbf{\bibinfo{volume}{48}},
  \bibinfo{pages}{102108} (\bibinfo{year}{2007}).

\bibitem[{\citenamefont{Dumitriu et~al.}(2007)\citenamefont{Dumitriu, Edelman,
  and Shuman}}]{dumitriu}
\bibinfo{author}{\bibfnamefont{I.}~\bibnamefont{Dumitriu}},
  \bibinfo{author}{\bibfnamefont{A.}~\bibnamefont{Edelman}}, \bibnamefont{and}
  \bibinfo{author}{\bibfnamefont{G.}~\bibnamefont{Shuman}},
  \bibinfo{journal}{J. Symb. Comp.} \textbf{\bibinfo{volume}{42}},
  \bibinfo{pages}{587} (\bibinfo{year}{2007}).

\bibitem[{\citenamefont{Dunkl and Xu}(2001)}]{dunkl2}
\bibinfo{author}{\bibfnamefont{C.}~\bibnamefont{Dunkl}} \bibnamefont{and}
  \bibinfo{author}{\bibfnamefont{Y.}~\bibnamefont{Xu}},
  \emph{\bibinfo{title}{Orthogonal Polynomials of Several Variables}}
  (\bibinfo{publisher}{Cambridge}, \bibinfo{address}{Cambridge},
  \bibinfo{year}{2001}).

\bibitem[{\citenamefont{Griffiths and Span{\`o}}(2011)}]{griffithsspano}
\bibinfo{author}{\bibfnamefont{R.~C.} \bibnamefont{Griffiths}}
  \bibnamefont{and}
  \bibinfo{author}{\bibfnamefont{D.}~\bibnamefont{Span{\`o}}},
  \bibinfo{journal}{Bernoulli} \textbf{\bibinfo{volume}{17}},
  \bibinfo{pages}{1095} (\bibinfo{year}{2011}).

\bibitem[{\citenamefont{Avron et~al.}(2007)\citenamefont{Avron, Bisker, and
  Kenneth}}]{avron}
\bibinfo{author}{\bibfnamefont{J.~E.} \bibnamefont{Avron}},
  \bibinfo{author}{\bibfnamefont{G.}~\bibnamefont{Bisker}}, \bibnamefont{and}
  \bibinfo{author}{\bibfnamefont{O.}~\bibnamefont{Kenneth}},
  \bibinfo{journal}{J. Math. Phys.} \textbf{\bibinfo{volume}{48}},
  \bibinfo{pages}{102107} (\bibinfo{year}{2007}).

\bibitem[{\citenamefont{Avron and Kenneth}(2009)}]{avron2}
\bibinfo{author}{\bibfnamefont{J.~E.} \bibnamefont{Avron}} \bibnamefont{and}
  \bibinfo{author}{\bibfnamefont{O.}~\bibnamefont{Kenneth}},
  \bibinfo{journal}{Ann. Phys.} \textbf{\bibinfo{volume}{324}},
  \bibinfo{pages}{470} (\bibinfo{year}{2009}).

\bibitem[{\citenamefont{Verstraete
  et~al.}(2001{\natexlab{c}})\citenamefont{Verstraete, Dehaene, and
  Moor}}]{ver8}
\bibinfo{author}{\bibfnamefont{F.}~\bibnamefont{Verstraete}},
  \bibinfo{author}{\bibfnamefont{J.}~\bibnamefont{Dehaene}}, \bibnamefont{and}
  \bibinfo{author}{\bibfnamefont{B.~D.} \bibnamefont{Moor}},
  \bibinfo{journal}{Phys. Rev. A} \textbf{\bibinfo{volume}{64}},
  \bibinfo{pages}{010101(R)} (\bibinfo{year}{2001}{\natexlab{c}}).

\bibitem[{\citenamefont{Leinaas et~al.}(2006)\citenamefont{Leinaas, Myrheim,
  and Ovrum}}]{leinaas}
\bibinfo{author}{\bibfnamefont{J.~M.} \bibnamefont{Leinaas}},
  \bibinfo{author}{\bibfnamefont{J.}~\bibnamefont{Myrheim}}, \bibnamefont{and}
  \bibinfo{author}{\bibfnamefont{E.}~\bibnamefont{Ovrum}},
  \bibinfo{journal}{Phys. Rev. A.} \textbf{\bibinfo{volume}{74}},
  \bibinfo{pages}{012313} (\bibinfo{year}{2006}).

\bibitem[{\citenamefont{Szarek et~al.}(2008)\citenamefont{Szarek, Werner, and
  {\.Z}yczkowski}}]{SWZ}
\bibinfo{author}{\bibfnamefont{S.}~\bibnamefont{Szarek}},
  \bibinfo{author}{\bibfnamefont{E.}~\bibnamefont{Werner}}, \bibnamefont{and}
  \bibinfo{author}{\bibfnamefont{K.}~\bibnamefont{{\.Z}yczkowski}},
  \bibinfo{journal}{J. Math. Phys.} \textbf{\bibinfo{volume}{49}},
  \bibinfo{pages}{032113} (\bibinfo{year}{2008}).

\bibitem[{\citenamefont{Szarek et~al.}(2011)\citenamefont{Szarek, Werner, and
  {\.Z}yczkowski}}]{SWZ2}
\bibinfo{author}{\bibfnamefont{S.}~\bibnamefont{Szarek}},
  \bibinfo{author}{\bibfnamefont{E.}~\bibnamefont{Werner}}, \bibnamefont{and}
  \bibinfo{author}{\bibfnamefont{K.}~\bibnamefont{{\.Z}yczkowski}},
  \bibinfo{journal}{J. Phys. A} \textbf{\bibinfo{volume}{44}},
  \bibinfo{pages}{045303} (\bibinfo{year}{2011}).

\bibitem[{\citenamefont{Lang and Caves}()}]{langcaves}
\bibinfo{author}{\bibfnamefont{M.~D.} \bibnamefont{Lang}} \bibnamefont{and}
  \bibinfo{author}{\bibfnamefont{C.~M.} \bibnamefont{Caves}},
  \eprint{arXiv:1006.2775}.

\bibitem[{\citenamefont{Simon and Taylor}(1985)}]{simontaylor}
\bibinfo{author}{\bibfnamefont{B.}~\bibnamefont{Simon}} \bibnamefont{and}
  \bibinfo{author}{\bibfnamefont{M.}~\bibnamefont{Taylor}},
  \bibinfo{journal}{Commun. Math. Phys.} \textbf{\bibinfo{volume}{101}},
  \bibinfo{pages}{1} (\bibinfo{year}{1985}).

\bibitem[{\citenamefont{Miller}(1977)}]{miller}
\bibinfo{author}{\bibfnamefont{W.}~\bibnamefont{Miller}, \bibfnamefont{Jr.}},
  \emph{\bibinfo{title}{Symmetry and Separation of Variables}}
  (\bibinfo{publisher}{Addison-Wesley}, \bibinfo{address}{Reading},
  \bibinfo{year}{1977}).

\bibitem[{\citenamefont{Bargmann}(1947)}]{bargmann}
\bibinfo{author}{\bibfnamefont{V.}~\bibnamefont{Bargmann}},
  \bibinfo{journal}{Ann. of Math.} \textbf{\bibinfo{volume}{48}},
  \bibinfo{pages}{568} (\bibinfo{year}{1947}).

\bibitem[{\citenamefont{Batle et~al.}(2002)\citenamefont{Batle, Plastino,
  Casas, and Plastino}}]{batleplastino1}
\bibinfo{author}{\bibfnamefont{J.}~\bibnamefont{Batle}},
  \bibinfo{author}{\bibfnamefont{A.~R.} \bibnamefont{Plastino}},
  \bibinfo{author}{\bibfnamefont{M.}~\bibnamefont{Casas}}, \bibnamefont{and}
  \bibinfo{author}{\bibfnamefont{A.}~\bibnamefont{Plastino}},
  \bibinfo{journal}{Phys. Lett. A} \textbf{\bibinfo{volume}{298}},
  \bibinfo{pages}{301} (\bibinfo{year}{2002}).

\bibitem[{\citenamefont{Kwok and Chen}(2000)}]{kwokchen}
\bibinfo{author}{\bibfnamefont{W.}~\bibnamefont{Kwok}} \bibnamefont{and}
  \bibinfo{author}{\bibfnamefont{Z.}~\bibnamefont{Chen}}, in
  \emph{\bibinfo{booktitle}{Proceedings of the Ninth International Meshing
  Round Table, New Orleans LA}} (\bibinfo{year}{2000}), pp.
  \bibinfo{pages}{325--333}.

\end{thebibliography}

\end{document}